\documentclass{article}
\usepackage{slashed,amsmath,amssymb,enumitem}
\usepackage[papersize={8.5in,11in}]{geometry}
\usepackage{slashed,feynmf,epsfig,verbatim,cite}

\begin{document}

\title{Radiative Corrections In An Ultraviolet Complete Electroweak Model Without A Higgs Particle}
\author{J. W. Moffat\\~\\
Perimeter Institute for Theoretical Physics,\\ Waterloo, Ontario N2L 2Y5, Canada\\
and\\
Department of Physics and Astronomy, University of Waterloo,\\ Waterloo, Ontario N2L 3G1, Canada}
\date{\today}

\maketitle

\begin{abstract}
The one-loop radiative correction in an ultraviolet (UV) complete electroweak (EW) model without a Higgs particle is calculated. The $\rho$ parameter determining the ratio of the charged to neutral currents is derived for the dominant top quark contribution with the result $\rho\sim 1.01$. This result is compared to the radiative correction to $\rho$ in the standard EW model with a Higgs particle. For the favored light Higgs particle mass from global fits to EW data: $114.4\,{\rm GeV} \leq m_H\leq 135\, {\rm GeV}$, the Higgs contribution up to three loops is negligible and the Higgless model is consistent with EW data for the energy scale $\Lambda_W\lesssim 1-2$ TeV.
\end{abstract}

\vskip 0.2 true in
e-mail: john.moffat@utoronto.ca, jmoffat@perimeterinstitute.ca



\begin{fmffile}{nhfigs}

\fmfcmd{%
 vardef bar (expr p, len, ang) =
  ((-len/2,0)--(len/2,0))
     rotated (ang + angle
       direction length(p)/2 of p)
       shifted point length(p)/2 of p
 enddef;
 style_def crossed_phantom expr p =
  draw_phantom p;
  ccutdraw bar (p, 3mm,  60);
  ccutdraw bar (p, 3mm, -60)
 enddef;}

\section{Introduction}

 An electroweak (EW) model has been developed based on a quantum field theory which is finite to all orders of perturbation theory~\cite{Moffat2010}. The broken $SU_L(2)\times U_Y(1)$ symmetry is treated as dynamically intrinsic to the model, retaining the $U_{\rm EM}(1)$ invariance for a massless photon. No attempt is made to explain the origin of the fermion and boson masses and the model does not contain a physical scalar field in the action and there is no Higgs particle. The model is finite to all orders of perturbation theory removing the need for renormalizability to produce finite calculations of cross sections. The unitarity of the $W_LW_L\rightarrow W_LW_L$ scattering amplitude is guaranteed by the suppression of unitarity violation by the running coupling constant $g(s)\lesssim 1/\sqrt{s}$ for the center-of-mass energy $\sqrt{s} > 1-2$ TeV. The minimal model contains the observed three families of quarks and leptons and the bosons $W^\pm, Z^0$ and the massless photon and gluon.

\section{The Electroweak Model}
\label{sec:local}

Our theory is based on the local $SU(3)_c\times SU_L(2)\times U_Y(1)$ Lagrangian that includes leptons and quarks with the color degree of freedom of the strong interaction group $SU_c(3)$. We shall use the metric convention, $\eta_{\mu\nu}={\rm diag}(+1,-1,-1,-1)$, and set $\hbar=c=1$. The EW sector of the model is based on the Lagrangian:
\begin{align}
\label{Lagrangian}
{\cal L}_{\rm EW}=\sum_{\psi_L}\bar\psi_L\biggl[\gamma^\mu\biggl(i\partial_\mu - {\bar g}T^aW^a_\mu - {\bar g}'\frac{Y}{2}B_\mu\biggr)\biggr]\psi_L\nonumber\\
+\sum_{\psi_R}\bar\psi_R\biggl[\gamma^\mu\biggl(i\partial_\mu - {\bar g}'\frac{Y}{2}B_\mu\biggr)\biggr]\psi_R -\frac{1}{4}B^{\mu\nu}B_{\mu\nu}\nonumber\\
-\frac{1}{4}W_{\mu\nu}^aW^{a\mu\nu} + {\cal L}_M + {\cal L}_{m_f}.
\end{align}
The fermion fields (leptons and quarks) have been written as $SU_L(2)$ doublets and $U_Y(1)$ singlets, and we have suppressed the fermion generation indices. We have $\psi_{L,R}=P_{L,R}\psi$, where $P_{L,R}=\frac{1}{2}(1\mp\gamma_5)$. Moreover,
\begin{equation}
\label{Bequation}
B_{\mu\nu}=\partial_\mu B_\nu-\partial_\nu B_\mu,
\end{equation}
and
\begin{equation}
W^a_{\mu\nu}=\partial_\mu W_\nu^a-\partial_\nu W_\mu^a-{\bar g}f^{abc}W_\mu^bW_\nu^c.
\end{equation}
The quark and lepton fields and the boson fields $W^a_\mu$ and $B_\mu$ are {\it local fields} that satisfy microcausality.

The ${\bar g}$ and ${\bar g}'$ are defined by
\begin{equation}
{\bar g}(x)=g{\cal E}(\Box(x)/2\Lambda_W^2),\quad {\bar g}'(x)=g'{\cal E}(\Box(x)/2\Lambda_W^2),
\end{equation}
where $\Lambda_W$ is an energy scale that is a measurable parameter in the model and ${\cal E}$ is a differential operator and an {\it entire function} of $\Box=\partial^\mu\partial_\mu$.

The Lagrangian for the boson mass terms is
\begin{equation}
{\cal L}_M=\frac{1}{2}M_{W^a}^2W^{a\mu} W^a_\mu + \frac{1}{2}M_B^2B^\mu B_\mu,
\end{equation}
and the fermion mass Lagrangian is
\begin{equation}
\label{fermionmass}
{\cal L}_{m_f}=-\sum_{\psi_L^i,\psi_R^j}m_{ij}^f(\bar\psi_L^i\psi_R^j + \bar\psi_R^i\psi_L^j),
\end{equation}
where $M_{W^a}, M_B$ and $m_{ij}^f$ denote the boson and fermion masses, respectively. Eq.(\ref{fermionmass}) can incorporate massive neutrinos and their flavor oscillations.

Both of these mass Lagrangians explicitly break $SU(2)_L\times U(1)_Y$ gauge symmetry. In the standard model the $SU_L(2)\times U_Y(1)$ symmetry of the vacuum is spontaneously broken and the non-zero, scalar field vacuum expectation value gives mass to the fermions through an $SU_L(2)\times U_Y(1)$ invariant Yukawa Lagrangian. However, this requires 12 coupling constant parameters, which are fixed to generate the observed masses of the 12 quarks and leptons, and a Higgs particle with a mass $m_H$ which is not predicted by the theory.

The $SU(2)$ generators satisfy the commutation relations
\begin{equation}
[T^a,T^b]=if^{abc}T^c,~~~~~\mathrm{with}~~~~~T^a=\frac{1}{2}\sigma^a.
\end{equation}
Here, $\sigma^a$ are the Pauli spin matrices and $f^{abc}=\epsilon^{abc}$. The fermion--gauge boson interaction terms are contained in
\begin{equation}
\label{interactLagr}
{\cal L}_I=-i{\bar g}J^{a\mu}W_\mu^a-i{\bar g}'J_Y^\mu B_\mu,
\end{equation}
where the $SU(2)$ and hypercharge currents are given by
\begin{equation}
J^{a\mu}=-i\sum_{\psi_L}\bar{\psi}_L\gamma^\mu T^a\psi_L,~~~~~\mathrm{and}~~~~~J_Y^\mu=-i\sum_\psi\frac{1}{2}Y\bar\psi\gamma^\mu\psi,
\end{equation}
respectively, and the nonlocal differential operator ${\cal E}(\Box/2\Lambda_W^2)$ acts on the right on the currents $J^{a\mu}(x)$ and $J^\mu_Y(x)$. The last sum is over all left and right-handed fermion states with hypercharge factors $Y=2(Q-T^3)$ where $Q$ is the electric charge.
We define for later notational convenience $\slashed W=\gamma^\mu W_\mu^aT^a$.

We can equally well write the interaction Lagrangian (\ref{interactLagr}) in the form:
\begin{equation}
{\cal L}_I=-ig{\cal J}^{a\mu}W_\mu^a-ig'{\cal J}_Y^\mu B_\mu,
\end{equation}
where
\begin{equation}
{\cal J}^{a\mu}=-i\sum_{\psi_L}\bar{\Psi}_L\gamma^\mu T^a\Psi_L,~~~~~\mathrm{and}~~~~~{\cal J}_Y^\mu=-i\sum_\Psi\frac{1}{2}Y\bar\Psi\gamma^\mu\Psi.
\end{equation}
Here,
\begin{equation}
{\bar\Psi}(x)={\cal E}(\Box(x)/4\Lambda_W^2){\bar\psi}(x),~~~~~\mathrm{and}~~~~~\Psi(x)={\cal E}(\Box(x)/4\Lambda_W^2)\psi(x),
\end{equation}
where $\bar\psi(x)$ and $\psi(x)$ denote the local fermion field operators.

We diagonalize the charged sector and perform mixing in the neutral boson sector. We write
$W^\pm=\frac{1}{\sqrt{2}}(W^1\mp iW^2)$ as the physical charged vector boson fields.
In the neutral sector, we can mix the fields in the usual way:
\begin{equation}
Z_\mu=\cos\theta_wW_\mu^3-\sin\theta_wB_\mu~~~\mathrm{and}~~~A_\mu=\cos\theta_wB_\mu+\sin\theta_wW_\mu^3.
\label{eq:2.35}
\end{equation}
We define the usual relations
\label{sintheta}
\begin{equation}
\sin^2\theta_w=\frac{g'^2}{g^2+g'^2}~~~\mathrm{and}~~~\cos^2\theta_w=\frac{g^2}{g^2+g'^2}.
\end{equation}

If we identify the resulting $A_\mu$ field with the photon, then we have the unification condition:
\begin{equation}
\label{unifcond}
{\bar e}(x)={\bar g}(x)\sin\theta_w={\bar g}'(x)\cos\theta_w,
\end{equation}
where
\begin{equation}
\label{bare}
{\bar e}(x)=e{\cal E}(\Box(x)/2\Lambda_W^2).
\end{equation}

We can consider the possibility that the unification condition (\ref{unifcond}) is broken at high energies. We then have
\begin{equation}
{\bar g}(x)\sin\theta_w={\bar g}'(x)\cos\theta_w,
\end{equation}
and
\begin{equation}
{\bar e}(x) =e{\cal G}(\Box(x)/2\Lambda_{EM}^2).
\end{equation}
Here, ${\cal G}(x)$ is an entire function and the photon sector energy scale $\Lambda_{EM} > \Lambda_W \sim 1-2$ TeV.

The electromagnetic and neutral currents are given by
\begin{equation}
J_\mathrm{em}^\mu=J^{3\mu}+J_Y^\mu,
\end{equation}
and
\begin{equation}
J_\mathrm{NC}^\mu=J^{3\mu}-\sin^2\theta_wJ_\mathrm{em}^\mu.
\end{equation}
The fermion-boson interaction terms are given by
\begin{equation}
{\cal L}_I=-\frac{\bar g}{\sqrt{2}}(J_\mu^+W^{+\mu}+J_\mu^-W^{-\mu})-{\bar g}\sin\theta_wJ_\mathrm{em}^\mu A_\mu-\frac{\bar g}{\cos\theta_w}
J_\mathrm{NC}^\mu Z_\mu.
\end{equation}

Gauge invariance is important for the QED sector, $U_{\rm em }(1)$, for it leads to a consistent quantization of QED calculations by guaranteeing that the Ward-Takahashi identities are valid~\cite{Moffat2010,Clayton1993,Moffat1991,Moffat2011}. The quantization of the Proca massive vector boson sector of $SU(2)\times U(1)$ is physically consistent even though the $SU(2)\times U(1)$ gauge symmetry is dynamically broken~\cite{Moffat2010}.

\section{The UV Complete EW Theory}
\label{sec:regularized}

To write the theory in its finite, nonlocal form, we follow the method outlined in~\cite{Moffat2010}.
The key observation in our nonlocal QFT is that the vertices contain nonlocal factors (as they do in string field theory, for instance), which causes loops to converge in Euclidean space. The nonlocal coupling form factor $g(x)$ at vertices has the effect of smearing fields in loop integrals, thereby avoiding the divergences of local point particle QFT. The momentum space Fourier transforms of the coupling constants ${\bar g}, {\bar g}'$ and ${\bar e}$ are {\it entire functions} in momentum space, which are complex differentiable and holomorphic everywhere in the complex plane, except at infinity. They do not introduce unphysical poles into the propagators of the theory guaranteeing unitarity. The nonlocal coupling function can be related to a Lorentz invariant operator distribution as
\begin{equation}
{\bar g}(x)=g{\cal E}(x-y,\Lambda_W)\delta^{(4)}(x-y),
\end{equation}
where $\Lambda_W$ denotes a nonlocal electroweak energy scale. A possible simple choice of a specific smearing operator is
\begin{equation}
{\cal E}(x,\Lambda_W)=\exp\left(-\frac{\Box+m^2}{2\Lambda_W^2}\right).
\end{equation}

We adopt an economical intrinsic breaking of $SU(2)\times U(1)$ symmetry by stipulating that the massive boson Lagrangian takes the form~\cite{Moffat2010}:
\begin{align}
\label{massmatrix}
{\cal L}_M=&\frac{1}{8}b^2g^2[(W^1_\mu)^2+(W^2_\mu)^2]+\frac{1}{8}b^2[g^2(W^3_\mu)^2-2gg'W^3_\mu B^\mu+g^{'2}B^2_\mu]\nonumber\\
=&\frac{1}{4}g^2b^2W^+_\mu W^{-\mu}+\frac{1}{8}b^2(W_{3\mu},B_\mu)\left(\begin{matrix}g^2&-gg'\\
-gg'&g^{'2}\end{matrix}\right)\left(\begin{matrix}W^{3\mu}\\B^\mu\end{matrix}\right),
\end{align}
where $b$ is the EW symmetry breaking energy scale. We have the usual symmetry breaking mass matrix in which one of the eigenvalues of the $2\times 2$ matrix in (\ref{massmatrix}) is zero, which leads to the mass values:
\begin{equation}
\label{bosonmasses}
M_W=\frac{1}{2}bg,\quad M_Z=\frac{1}{2}b(g^2+g^{\prime 2})^{1/2},\quad M_A=0.
\end{equation}
We do not identify $b$ with the vacuum expectation value $v=\langle\phi\rangle_0$ in the standard Higgs model. The boson mass Lagrangian is given by
\begin{equation}
{\cal L}_M=M_W^2W^+_\mu W^{-\mu} + \frac{1}{2}M_Z^2Z_\mu Z^\mu.
\end{equation}
We do not know the origin of the symmetry breaking mechanism and scale $b$. Postulating the dynamical breaking of the $SU(2)\times U(1)$ symmetry is no worse than adopting the {\it ad hoc} assumption of an invariant Lagrangian for the scalar fields when motivating the Higgs mechanism:
\begin{equation}
\label{scalarLagrangian}
{\cal L}_\phi=\large\vert(i\partial_\mu - gT^aW^a_\mu - g'\frac{Y}{2}B_\mu)\phi\large\vert^2-V(\phi),
\end{equation}
where $\vert...\vert^2=(...)^\dagger(...)$,
\begin{equation}
\label{potential}
V(\phi)=\mu^2\phi^\dagger\phi +\lambda(\phi^\dagger\phi)^2,
\end{equation}
and $\mu^2 < 0$, $\lambda > 0$. There is no known fundamental motivation for choosing $\mu^2 < 0$ and we could add an additional contribution $\lambda'\phi^6$ to the potential (\ref{potential}) or even higher order polynomials in $\phi$. Such higher-dimensional operators will render the standard EW model with local, point-like interactions non-renormalizable. The quark and lepton masses and the $W$ and $Z$ masses are the physical masses in the propagators. We circumvent the problem of the lack of renormalizability of our model by damping out divergences with the coupling vertices ${\bar g}(p^2)$, ${\bar g}'(p^2)$ and ${\bar e}(p^2)$ in momentum space. We emphasize that our energy scale parameter $\Lambda_W \gtrsim 1$ TeV is not a naive cutoff. The entire function feature of the coupling vertices guarantees that the model suffers no violation of unitarity or Poincar\'e invariance.

From the relation
\begin{equation}
\label{bscale}
\frac{1}{2b^2}=\frac{g^2}{8M_W^2}=\frac{G_F}{\sqrt{2}},
\end{equation}
where $G_F=1.166\times 10^{-5}\,{\rm GeV}^{-2}$ is Fermi's constant determined from muon decay, we obtain the EW energy scale $b\sim 246$ GeV. The choice of intrinsic symmetry breaking (\ref{massmatrix}) guarantees that at the effective tree graph level:
\begin{equation}
\label{rhotree}
\rho^{\rm tree}=\frac{M_W^2}{M_Z^2\cos^2\theta_w}=1.
\end{equation}

To retain the $U_{\rm EM}(1)$ gauge invariance of the QED sector, we adopt a generalization of local gauge invariance based on a nonlocal gauge transformation~\cite{Moffat1991,Moffat2010,Clayton1993}. In practice we restore gauge invariance and ensure that the tree graphs remain local and point-like by introducing additional interaction terms into the Lagrangian, enforcing decoupling of unphysical degrees of freedom. This procedure can be repeated, order by order, and as shown in \cite{Moffat1991}\footnote{In \cite{Moffat1991}, the authors concern themselves primarily with QED; however, in the last section of the paper, the non-Abelian case is discussed and proof is offered that there is at least one solution to all orders.}, the exact form to all orders of the nonlocal gauge invariant QED is derived. Current conservation and the Ward-Takahashi identities for the nonlocal symmetry transformations have been derived to higher orders~\cite{Clayton1993}.

We observe that this theory is only rigorously defined in Euclidean space, but since it has been shown that an analytic continuation to Minkowski space always exists via Efimov's regulator \cite{Efimov1972,Moffat2010}, we will work in Minkowski space, only referring to Euclidean space to ensure the convergence of the loop integrals.

The quantization of the $SU(2)\times U(1)$ symmetry broken sector has been performed for the boson Proca action and it was shown how the longitudinal scalar components $W^a_0$ do not generate unphysical ghosts even in the absence of an explicit gauge invariance~\cite{Moffat2010}. Quantization can also be accomplished via the path integral formalism, which gives the expectation value of operators ${\cal O}$ as a sum over all field configurations weighted by the exponential of the classical action:
\begin{equation}
\label{expectationvalue}
\left<T({\cal O}[\phi])\right>\propto\int[D\bar\psi][D\psi][DW][DB]\mu[\bar\psi,\psi,B,W]{\cal O}[\phi]
\exp\left(i\int d^4x{\cal L}_\mathrm{EW}\right).
\end{equation}
To maintain gauge invariance and BRST invariance, ghost fields have to be included in the Lagrangian~\cite{Moffat1991}.

To generate a perturbation scheme in the field operators, we write the generating functional as
\begin{align}
W[{\cal J}]=&\ln(Z[{\cal J}])\nonumber\\
=&\ln\left(\int[D\phi]\exp\left(i\int dx\{L_F[\phi]+{\cal L}_I[{\bar g},{\bar g}',J,\phi]
+{\cal J}(x)\phi(x)\}\right)\right),
\end{align}
where the source term ${\cal J}$ is nonlocal and $\phi$ denotes the EW field operators ${\bar\psi},\psi, W, B, A$. We have
\begin{equation}
Z[\phi,{\cal J}]=\exp\biggl(i\int d^4x{\cal L}_I[\phi]\biggl[\frac{1}{i}\frac{\delta}{\delta{\cal J}(x)}\biggr]Z_0[\phi,{\cal J}]\biggr)
\end{equation}
and
\begin{equation}
Z_0[\phi,{\cal J}]\propto \exp\biggl[\frac{1}{2}\int d^4xd^4y(i{\cal J}(x)iD(x-y)i{\cal J}(y))\biggr].
\end{equation}
Green's functions and the Feynman rules are generated by
\begin{equation}
\langle 0\vert T[\phi(x_1)...\phi(x_n)]\vert 0\rangle=i^n\frac{\delta^nW[{\cal J}]}
{\delta{\cal J}(x_1)...\delta{\cal J}(x_n)}.
\end{equation}

The bare boson propagator in a general gauge is given by
\begin{equation}
iD^{\mu\nu}(p^2)=i\biggl(\frac{\eta^{\mu\nu}-\frac{p^\mu p^\nu}{p^2}}{p^2-M_{0V}^2+i\epsilon}
+\frac{\xi\frac{p^\mu p^\nu}{p^2}}{p^2-\xi M^2_{0V}+i\epsilon}\biggr),
\end{equation}
where $\xi$ is the gauge parameter. The fermion propagator is
\begin{equation}
iS(p)=\frac{i}{\slashed p-m_{0f}+i\epsilon},
\end{equation}
and $M_{0V}$ and $m_{0f}$ are the bare vector boson and fermion masses, respectively.

Quantizing the theory described by (\ref{expectationvalue}) in the path integral formalism requires finding
a measure that respects the full nonlocal nature of the action. We therefore require a method to generate
consistency conditions on a measure to retain the nonlocal quantum regime. The simplest way to do this
is to consider how the trivial measure transforms under the nonlocal gauge
transformations, and require that there is a contribution from the measure that compensates.

We write the full invariant measure as the product of the trivial measure and an exponentiated
action term~\cite{Moffat1993}:
\begin{equation}
\mu_{\rm inv}[\phi]=d\phi\exp(iS_{\rm meas}[\phi]),
\end{equation}
and then perform a gauge transformation and require that the full measure be invariant.
Functionally integrating to derive the measure yields:
\begin{equation}
\delta\mu_{\rm inv}[\phi]=\mu_{\rm inv}[\phi]\biggl(i\delta S_{\rm meas}+{\rm Tr}\biggl[\frac{\partial}{\partial\phi}\delta\phi\biggr]\biggr)=0.
\end{equation}
The trace appears as the only surviving diagonal terms of the Jacobian determinant of
the infinitesimal transformation (when dealing with fermions, the grassman derivatives will
produce the necessary extra minus sign that corresponds to the inverse determinant).
The condition to be satisfied by an invariant measure is
\begin{equation}
\delta S_{\rm meas}=i{\rm Tr}\biggl[\frac{\partial}{\partial\phi}\delta\phi\biggr].
\end{equation}
The measure is also constrained to be an entire function of the 4-momentum invariants for the particular process, ensuring that no additional unphysical degrees of freedom become excited in the quantum regime.

To produce the necessary contribution to vacuum polarization in order to satisfy the Ward identity and keep the photon transverse to lowest order:
\begin{equation}
S_{\rm meas}=-\frac{\Lambda_W^2}{4\pi^2}\int dpdq\delta^4(p+q)A^\mu(q)\Upsilon_{\mu\nu}(q)A^\nu(q),
\end{equation}
where
\begin{equation}
\Upsilon_{\mu\nu}(q)=\eta_{\mu\nu}\int^{1/2}_0dt(1-t)\exp\biggl(t\frac{q^2}{\Lambda_W^2}-\frac{1}{1-t}
\frac{m_f^2}{\Lambda_W^2}\biggr).
\end{equation}

This procedure determines the measure up to arbitrary gauge invariant terms but there is a natural minimal choice determined through relating the measure to the loop graph it corrects, resulting in a unique (if it exists) measure. Any other invariant terms properly belong in the action and should not be introduced into the measure.

\section{Calculation of EW Vacuum Polarization}
\label{sec:measure}

To calculate radiative decays in the quantum regime of our EW model, we break $SU_L(2)\times U_Y(1)$ down to $U_\mathrm{em}(1)$ by choosing a symmetry breaking measure $\mu_{\rm SB}$ in the path integral, so that the effects show up at loop order. This means choosing a measure which alters the quantization of the theory, in order to produce the desired results. Even though the choice of the symmetry breaking measure is not unique, after an initial {\it ansatz} chosen as the minimal scheme, the rest of the method follows directly. For the broken $SU(2)\times U(1)$ symmetry, we can obtain a measure $\mu_{\rm SB}$ that contributes a measure diagram in the calculation of loop graphs, which guarantees that the corrections to $\rho^{\rm tree}=1$ are small in agreement with experimental EW data.

The symmetry breaking measure in our path integral generates three scalar degrees of freedom that give the $W^\pm$ and $Z^0$ longitudinal modes associated with the three degrees of freedom of massive vector bosons, while retaining a massless photon.
The vacuum polarization tensor is defined by
\begin{equation}
\Pi^{\mu\nu}(p^2)=\biggl(\eta^{\mu\nu}-\frac{p^\mu p^\nu}{p^2}\biggr)\Pi^{T\mu\nu}+\frac{p^\mu p^\nu}{p^2}\Pi^{L\mu\nu}.
\end{equation}

Since we want to mix the $W^3$ and $B$ to get a massive $Z$ and a massless photon, we need to work with the measure in a sector which is common to all gauge bosons. This implies working with the fermion contributions and leaving the bosonic and ghost contributions invariant. We take it as given that the fermions have a mass. In the diagonalized $W^\pm$ sector, we get
\begin{align}
-i\Pi_{W^\pm}^L=&-\frac{ig^2\Lambda_W^2}{(4\pi)^2}\sum_{q^L}(K_{m_1m_2}-L_{m_1m_2}),\\
-i\Pi_{W^\pm}^T=&-\frac{ig^2\Lambda_W^2}{(4\pi)^2}\sum_{q^L}(K_{m_1m_2}-L_{m_1m_2}+2P_{m_1m_2}),
\label{eq:WWPi}
\end{align}
where we define
\begin{align}
K_{m_1m_2}=&\int_0^{\frac{1}{2}}d\tau(1-\tau)\left[\exp\left(-\tau\frac{p_E^2}{\Lambda_W^2}-f_{m_1m_2}\right)+
\exp\left(-\tau\frac{p_E^2}{\Lambda_W^2}-f_{m_2m_1}\right)\right],
\end{align}
\begin{align}
\label{eq:P}P_{m_1m_2}=&-\frac{p_E^2}{\Lambda_W^2}\int_0^{\frac{1}{2}}d\tau\tau(1-\tau)\left[E_1\left(\tau\frac{p_E^2}
{\Lambda_W^2}+f_{m_1m_2}\right)+E_1\left(\tau\frac{p_E^2}{\Lambda_W^2}+f_{m_2m_1}\right)\right],
\end{align}
\begin{align}
L_{m_1m_2}=&\int_0^{\frac{1}{2}}d\tau(1-\tau)\left[f_{m_1m_2}E_1\left(\tau\frac{p_E^2}{\Lambda_W^2}+f_{m_1m_2}\right)
+f_{m_2m_1}E_1\left(\tau\frac{p_E^2}{\Lambda_W^2}+f_{m_2m_1}\right)\right],
\end{align}
\begin{align}
M_{m_1m_2}=&\frac{m_1m_2}{\Lambda_W^2}\int_0^{\frac{1}{2}}d\tau\left[E_1\left(\tau\frac{p_E^2}{\Lambda_W^2}
+f_{m_1m_2}\right)+E_1\left(\tau\frac{p_E^2}{\Lambda_W^2}+f_{m_2m_1}\right)\right],
\end{align}
and where
\begin{equation}
f_{m_1m_2}=\frac{m_1^2}{\Lambda_W^2}+\frac{\tau}{1-\tau}\frac{m_2^2}{\Lambda_W^2}.
\end{equation}
Here, $p_E$ denotes the Euclidean momentum and $E_1(z)$ is the exponential integral:
\begin{equation}
E_1(z)=\int_z^\infty dt\exp(-t)t^{-1}=-\ln(z)-\gamma-\sum_{n=1}^\infty \frac{(-z)^n}{nn!},
\end{equation}
where $\gamma$ is the Euler-Masheroni constant. When the longitudinal piece is nonzero in the unitary gauge (where only the physical particle spectrum remains), we have no unphysical poles in the longitudinal sector. In this way, we can assure ourselves that we are not introducing spurious degrees of freedom into the theory.

We note that at $p^2=0$:
\begin{equation}
\label{PiW}
-i\Pi_{W^\pm}^L\bigg|_{p^2=0}=-i\Pi_{W^\pm}^T\bigg|_{p^2=0}=-\frac{ig^2\Lambda_W^2}{(4\pi)^2}\sum_{q^L}(K_{m_1m_2}-L_{m_1m_2})
\bigg|_{p^2=0}\ne 0.
\end{equation}

We go on to calculate the self-energy in the $W^3$ sector as
\begin{align}
-i\Pi_{W^3W^3}^L=&-\frac{1}{2}\frac{ig^2\Lambda_W^2}{(4\pi)^2}\sum_\psi(K_{mm}-L_{mm}),\\
-i\Pi_{W^3W^3}^T=&-\frac{1}{2}\frac{ig^2\Lambda_W^2}{(4\pi)^2}\sum_\psi(K_{mm}-L_{mm}+2P_{mm}).
\end{align}
It is clear that if we want the $B$ sector to mix with this, we need to make the vacuum polarization tensor look very similar. This is what motivates the choice of symmetry breaking measure, after one makes the initial
{\it ansatz}. In the $B$ sector we have
\begin{align}
-i\Pi_{BB}^L=&-\frac{1}{2}\frac{ig'^2\Lambda_W^2}{(4\pi)^2}\sum_\psi[16(Q-T^3)^2(K_{mm}-L_{mm})+32Q(Q-T^3)M_{mm}],\\
-i\Pi_{BB}^T=&-\frac{1}{2}\frac{ig'^2\Lambda_W^2}{(4\pi)^2}\sum_\psi[16(Q-T^3)^2(K_{mm}-L_{mm}+2P_{mm})+32Q(Q-T^3)M_{mm}].
\end{align}
We write the Feynman diagram measure contribution as
\begin{equation}
\Upsilon_{\mu\nu}^{BB}=-\frac{ig'^2\Lambda_W^2}{(4\pi)^2}\eta_{\mu\nu}\sum_\psi\left[\left(\frac{1}{2}-8(Q-T^3)^2
\right)(K_{mm}-L_{mm})-16Q(Q-T^3)M_{mm}\right]
\end{equation}
and we are then left with
\begin{eqnarray}
-i\Pi_{BB}^L&=&-\frac{1}{2}\frac{ig'^2\Lambda_W^2}{(4\pi)^2}\sum_\psi(K_{mm}-L_{mm}),\\
-i\Pi_{BB}^T&=&-\frac{1}{2}\frac{ig'^2\Lambda_W^2}{(4\pi)^2}\sum_\psi[(K_{mm}-L_{mm})+32(Q-T^3)^2P_{mm}].
\end{eqnarray}
Note that the pieces that contribute to the boson masses are identical to those given above. The presence of the extra piece proportional to $p^2$ will not give any problems in the mass matrix, and will produce a $Z$-photon mixing that contains no extra poles. The $B-W^3$ mixing sector originally looks like
\begin{align}
-i\Pi_{W^3B}^L=&-\frac{4igg'\Lambda_W^2}{(4\pi)^2}\sum_\psi[T^3(Q-T^3)(K_{mm}-L_{mm})+QM_{mm}],\\
-i\Pi_{W^3B}^T=&-\frac{4igg'\Lambda_W^2}{(4\pi)^2}\sum_\psi[T^3(Q-T^3)(K_{mm}-L_{mm}+2P_{mm})+QM_{mm}].
\end{align}
Thus, to make the mass contributions look identical, we write
\begin{equation}
\Upsilon_{\mu\nu}^{W^3B}=-\frac{igg'\Lambda_W^2}{(4\pi)^2}\eta_{\mu\nu}\sum_\psi\left[\left(-\frac{1}{2}-4T^3(Q-T^3)\right)
(K_{mm}-L_{mm})-4QM_{mm}\right].
\end{equation}
Then we have
\begin{eqnarray}
-i\Pi_{W^3B}^L&=&\frac{1}{2}\frac{igg'\Lambda_W^2}{(4\pi)^2}\sum_\psi(K_{mm}-L_{mm}),\\
-i\Pi_{W^3B}^T&=&\frac{1}{2}\frac{igg'\Lambda_W^2}{(4\pi)^2}\sum_\psi[(K_{mm}-L_{mm})-8T^3(Q-T^3)P_{mm}].
\end{eqnarray}

The diagonal $Z-Z$ piece has a longitudinal part
\begin{equation}
-i\Pi_{ZZ}^L=-\frac{1}{2}\frac{i(g^2+g'^2)\Lambda_W^2}{(4\pi)^2}\sum_\psi(K_{mm}-L_{mm}).
\end{equation}
In the transverse sector things look a bit more complicated. For the $Z-Z$ part we get
\begin{align}
-i\Pi_{ZZ}^T=&-\frac{1}{2}\frac{i(g^2+g'^2)\Lambda_W^2}{(4\pi)^2}\nonumber\\
&\times\sum_\psi[(K_{mm}-L_{mm})+P_{mm}(2c_w^4+s_w^432(Q-T^3)^2-16s_w^2c_w^2T^3(Q-T^3))],
\label{ZZPi}
\end{align}
where $c_w=\cos\theta_w$ and $s_w=\sin\theta_w$. The pure photon sector gives
\begin{align}
-i\Pi_{AA}^T=&-\frac{1}{2}\frac{i(g^2+g'^2)\Lambda_W^2}{(4\pi)^2}c_w^2s_w^2\nonumber\\
&\times\sum_\psi P_{mm}(2+32(Q-T^3)^2+16T^3(Q-T^3)).
\label{eq:photon}
\end{align}
We observe from (\ref{eq:photon}) that $\Pi_{AA}^T(0)=0$, as follows from (\ref{eq:P}), guaranteeing a massless photon.

Finally we obtain for the mixing sector:
\begin{align}
-i\Pi_{AZ}^T=&-\frac{1}{2}\frac{i(g^2+g'^2)\Lambda_W^2}{(4\pi)^2}c_w^2s_w^2\nonumber\\
&\times\sum_\psi P_{mm}[2c_w^2-32s_w^2(Q-T^3)^2-16T^3(Q-T^3)(s_w^2-c_w^2)].
\end{align}

\section{Calculation of Radiative Corrections in the UV Complete Model}
\label{sec:rho}

We can relate the {\it finite} renormalized self-energies to physical observables by using the $S, T$ and $U$ parameters developed to describe radiative corrections in EW precision data investigations~\cite{Takeuchi1990,Holdom1990,Altarelli1991}. The parameters are defined in terms of the transverse gauge boson self-energies at zero momentum and their first derivatives:
\begin{equation}
\Pi^T_X(0)=\Pi^T_X(p^2)\vert_{p^2=0},\quad \Pi^{T'}_{X}(0)=\frac{\partial}{\partial p^2}\Pi^T_X(p^2)\vert_{p^2=0},
\end{equation}
where $X=AA,AZ,ZZ,WW$. The self-energies $\Pi^T$ are not individually observable, and in the standard local EW theory they contain UV divergences. However, the combinations:
\begin{eqnarray}
S&=&\frac{4s_w^2c_w^2}{\alpha}\biggl(\Pi^{T'}_{ZZ}-\frac{c_w^2-s_w^2}{c_ws_w}\Pi^{T'}_{AZ}-\Pi^{T'}_{AA}\biggr),\\
T&=&\frac{1}{\alpha m_W^2}(c_w^2\Pi^T_{ZZ}-\Pi^T_{WW}),\\
U&=&\frac{4s_w^2}{\alpha}(\Pi^{T'}_{WW}-c_w^2\Pi^{T'}_{ZZ}-2c_ws_w\Pi^{T'}_{AZ}-s_w^2\Pi^{T'}_{AA}),
\end{eqnarray}
are observable.

We wish to compare the effects of radiative corrections in the EW Higgs model with our model that does not have a Higgs particle. For the light Higgs boson favored by global fits to EW data and for the minimal Higgs model, only the oblique radiative corrections are important~\cite{Moffat2010}. The non-oblique Higgs particle correction to the decay $Z\rightarrow {\bar b}b$ in the minimal model is negligible due to the small coupling of the $b$ quark to the $Z$. We make for a light Higgs the assumption that masses of external fermions in diagrams are small compared to $M_W$. Thus, any contributions coming from the longitudinal components of the boson self-energies are suppressed by $m_f^2/M_W^2$. We also assume that only constant $p^2=0$ and linear terms in $p^2$ are important, for higher order terms are suppressed by heavy particles.

We stress that the $S, T$ and $U$ parameters were defined to determine bounds on {\it new physics} i.e., the effects of exotic new particles with heavy masses $M_{\rm new} >> M_Z$ in an expansion in $M_Z/M_{\rm new}$ on gauge boson self-energies. Thus, corrections from the new physics to $\rho=M_W^2/M_Z^2c^2_w$ are exactly zero in the minimal standard Higgs model as well as in our model without a Higgs particle. It follows that in our model $S^{\rm new}=T^{\rm new}=U^{\rm new}=0$ and any contributions to these parameters {\it do not include any contributions from $m_t$ or $M_H$, which are treated separately in what follows.} The extension of the standard model from new exotic physics is described by the $\rho^0$ parameter:
\begin{equation}
\rho^0=\frac{M_W^2}{M_Z^2c^2_w\hat\rho},
\end{equation}
where $\hat\rho$ accounts for the radiative corrections due to $m_t$ and $M_H$. The $\rho^0\neq 1$ effects describe new sources of custodial $SU(2)$ breaking that cannot be accounted for by the Higgs doublet or $m_t$. There are sufficient data to determine $\rho^0, M_H, m_t$ and $\alpha_s$. Global fits give~\cite{PDG}:
\begin{equation}
\label{exotic}
\rho^0=1.0004^{+0.0008}_{-0.0004},\quad 114.4\,{\rm GeV} \leq M_H\leq 215\,{\rm GeV},
\end{equation}
and
\begin{equation}
m_t=171.2\pm 1.9\,{\rm GeV},\quad \alpha_s(M_Z)=0.1215\pm 0.0017.
\end{equation}
Ignoring the direct LEP search bound $114.4$ GeV, the global fits yield
\begin{equation}
M_H=76^{+111}_{-38}\,{\rm GeV},\quad \rho^0=1.0000^{+0.0011}_{-0.0007}.
\end{equation}
The result (\ref{exotic}) for $\rho^0$ is slightly above but consistent with the standard model and our model without a Higgs particle.

The perturbation theory requires a {\it finite} mass renormalization. For the vector bosons, we have
\begin{equation}
\label{expmass}
M_V^2=M_{0V}^2+\delta M_V^2,
\end{equation}
where $M_{0V}$ and $\delta M_V$ denote the $W$ and $Z$ bare and self-energy masses, respectively. We have
\begin{equation}
\label{eq:selfvecmass}
\delta M_V^2=\Pi_V^T(M_V^2).
\end{equation}

The physical $W$ mass $M_W$ can be determined by the position of the pole in the $W$ propagator:
\begin{equation}
D^T_{W\mu\nu}=\frac{\eta_{\mu\nu}}{p^2-M_{0W}^2-\Pi^T_{WW}(p^2)}.
\end{equation}
By expanding $\Pi^T_{WW}(p^2)$ to first order in $p^2$, we obtain the equation:
\begin{equation}
\label{massequation}
p^2-M_{0W}^2-\Pi^T_{WW}-p^2\Pi^{T'}_{WW}=0,
\end{equation}
where (\ref{massequation}) is satisfied on the mass shell: $p^2-M_{W^2,phys}=0$. It follows that
\begin{equation}
M_{W,phys}^2=\frac{M_{0W}^2+\Pi^T_{WW}}{1-\Pi^{T'}_{WW}}.
\end{equation}
For the physical $Z$ mass we get
\begin{equation}
M_{Z,phys}=\frac{M_{0Z}^2+\Pi^T_{ZZ}}{1-\Pi^{T'}_{ZZ}}.
\end{equation}

There will be a renormalization of the muon tree-level decay rate. From the low-energy Fermi model, we obtain
\begin{equation}
\frac{G_F}{\sqrt{2}}=\frac{e^2_0}{8s_W^2M_{0W}}\biggl(1+\frac{\Pi^T_{WW}}{M_W^2}\biggr).
\end{equation}
The self-energy corrections to the bare electric charge $e_0$ are give by
\begin{equation}
e_{phys}^2=e_0^2(1+\Pi^{T'}_{AA}).
\end{equation}

For the $Z$-boson, the on-shell mass $M_Z$ is well known from experiment. The right-hand side of (\ref{eq:selfvecmass}) for the $Z$ boson is determined by (\ref{ZZPi}), and we find that it contains terms that include the electroweak coupling constant, the Weinberg angle, fermion masses, and the $\Lambda_W$ parameter. As all these except $\Lambda_W$ and the bare mass $M_{0Z}$ are known from experiment, the equation
\begin{equation}
\delta M_Z^2=\Pi^T_Z(M_Z^2),
\end{equation}
the right-hand side of which contains $\Lambda_W$ through (\ref{ZZPi}), can be used to determine $\Lambda_W$ given the unobservable bare mass $M_{0Z}$. Let us assume that $M_{0Z}=0$ and using the values
\begin{eqnarray}
g&=&0.649,\\
\sin^2\theta_w&=&0.2312,\\
m_t&=&171.2~\mathrm{GeV},
\end{eqnarray}
we get
\begin{equation}
\Lambda_W=541.9~\mathrm{GeV},
\end{equation}
where the precision of $\Lambda_W$ is determined by the precision to which the $Z$-mass is known, $M_Z=91.1876\pm 0.0021$~GeV \cite{PDG}, and it is not sensitive to the lack of precision knowledge of the top quark mass or the other quark masses. Knowing $\Lambda_W$ allows us to solve the consistency equation for the $W$ boson mass. Treating $M_{0W}$ as unknown, we solve for $\delta M_W$ using (\ref{eq:WWPi}):
\begin{equation}
\delta M_W^2=\Pi^T_W(M_W^2),
\end{equation}
and obtain at the $W$ pole
\begin{equation}
\delta M_W^{(1)}\simeq 80.05~\mathrm{GeV}.
\end{equation}

We can also obtain a solution for the boson self-energies for $\Lambda_W=1$ TeV. Again the top quark dominates the calculations and we obtain
\begin{equation}
\label{biggerPi}
\Pi^T_{ZZ}(0)=(175.07\,{\rm GeV})^2,~~~\mathrm{and}~~~\Pi^T_{WW}(0)=(153.31\,{\rm GeV})^2.
\end{equation}
At the $Z$ and $W$ poles we get
\begin{equation}
\label{Pipoles}
\delta M_Z^2\equiv\Pi^T_Z(M_Z^2)=(170.14\,{\rm GeV})^2,~~~\mathrm{and}~~~\delta M_W^2\equiv\Pi^T_W(M_W^2)=(149.73\,{\rm GeV})^2.
\end{equation}

\section{The $\rho$ Parameter and its Radiative Corrections}

The EW $\rho$ parameter is a measure of the relative strengths of neutral and charged current interactions in four-fermion processes. At zero momentum transfer:
\begin{equation}
\label{radrho}
\rho=\frac{J_{NC}(0)}{J_{CC}(0)}=\frac{1}{1-\Delta\rho},
\end{equation}
where $J_{CC}(0)$ is given by the Fermi coupling constant $G_F$ determined from the $\mu$ decay rate, while $J_{NC}(0)$ is measured by neutrino scattering on electrons or hadrons and $\Delta\rho$ denotes radiative corrections to $\rho$. We have
\begin{equation}
{\cal L}^{NC}=-\frac{\rho G_F}{2\sqrt{2}}[{\bar \nu}_\mu\gamma^\mu(1-\gamma_5)\nu_\mu][{\bar e}\gamma_\mu(1-4s_W^2-\gamma_5)e],
\end{equation}
and
\begin{equation}
{\cal L}^{CC}=-\frac{G_F}{\sqrt{2}}[{\bar \nu}_\mu\gamma^\mu(1-\gamma_5)\mu][{\bar e}\gamma_\mu(1-4s_w^2-\gamma_5)\nu_e].
\end{equation}

In the minimal standard EW model, the spontaneous symmetry breaking and mass generation, based on the minimal assumption that the Higgs scalar field transforms as an isospin doublet, yields the tree-level relation between the weak angle $\theta_w$ and the $W$ and $Z$ boson masses given by (\ref{rhotree}).

In our EW model without a Higgs sector, we have postulated the intrinsic dynamical symmetry breaking (\ref{massmatrix}) and the effective tree level graphs obey (\ref{rhotree}). The tree-level relation gets modified by radiative corrections which are normally parameterized by (\ref{radrho})~\cite{Veltman1977}, where $\Delta\rho$ is dominated by the top quark. The dominant contributions come from the $W$ and $Z$ self-energies involving the $t$ and $b$ quark loops and they are proportional to $m_t^{2L}$, where $L$ is the number of loops.

In principle, radiative corrections to four-fermion processes include self-energy corrections to the boson propagators, vertex and box corrections, which can all affect $\rho$. It is possible to choose a gauge in which vertices and boxes do not contribute to $\Delta\rho$ at the leading order and only corrections coming from the transverse $W$ and $Z$ self-energies contribute. We have
\begin{equation}
\label{deltarho}
\Delta\rho=\frac{\Pi^T_{ZZ}(0)}{M_Z^2}-\frac{\Pi^T_{WW}(0)}{M_W^2}.
\end{equation}

Let us evaluate (\ref{deltarho}) in our EW model. We obtain from (\ref{PiW}) and (\ref{ZZPi}):
\begin{equation}
\Pi^T_{ZZ}(0)=8791.9~{\rm GeV}^2,\quad \Pi^T_{WW}(0)=6702.02~{\rm GeV}^2.
\end{equation}
We use the calculation of $M_W$ and $M_Z$ obtained from our self-consistent calculation of $\Lambda_W=541.9$ GeV:
\begin{equation}
M_Z=91.1876\pm 0.0021~\mathrm{GeV},\quad M_W\sim\delta M_W\simeq 80.05~\mathrm{GeV}
\end{equation}
and using (\ref{deltarho}) we obtain
\begin{equation}
\Delta\rho^{(1)}(0)=0.0115.
\end{equation}

We prefer to have a larger energy scale, $\Lambda_W=1$ TeV. We now adopt non-zero bare $W$ and $Z$ masses, $M_{0V}\neq 0$, and by using (\ref{biggerPi}) we obtain
\begin{equation}
\delta\rho^{(1)}(0)=0.0498.
\end{equation}
However, the calculation of $\delta\rho^{(1)}$ at the $Z$ and $W$ poles gives
\begin{equation}
\delta\rho^{(1)}=\frac{\Pi^T_{ZZ}(M_Z^2)}{M_Z^2}-\frac{\Pi^T_{WW}(M_W^2)}{M_W^2}=0.0130.
\end{equation}
This is in reasonable agreement with the value $\Delta\rho_t=0.01015$ obtained for the top quark dominated radiative correction~\cite{PDG}. We have not included the pure $W$ and $Z$ bosonic loop corrections in our estimations of $\Delta\rho$.

\section{Radiative Corrections in EW models}

We should investigate to what extent we can consider that our EW model without a Higgs particle and without a spontaneous symmetry breaking phase can fit the EW data. A lower bound on the Higgs mass $m_H > 114.4$ GeV has been established by direct searches at the LEP accelerator~\cite{LEP}. The EW precision data are sensitive to $m_H$ through quantum corrections and yield the range~\cite{EWworkgroups}:
\begin{equation}
m_H=87^{+35}_{-26}\, {\rm GeV}.
\end{equation}
The upper limit of the Higgs mass is $157$ GeV to the $95\%$ confidence-level based on using the EW data, or $186$ GeV if the LEP direct lower limit is included. The Tevatron experiments CDF and D0 have excluded the range~\cite{Tevatron}:
\begin{equation}
158\,{\rm GeV} < m_H < 175\,{\rm GeV}.
\end{equation}
Fitting all the data, yields the result
\begin{equation}
m_H=116.4^{+15.6}_{-1.3}\, {\rm GeV},
\end{equation}
at the $68\%$ confidence level.

At the tree level, we get $\rho=1$ due to the intrinsic symmetry breaking of $SU(2)\times U(1)$. This is due to the so-called custodial symmetry $SU_L(2)\times SU_R(2)$, which originates in the standard model from spontaneous symmetry breaking. We do not assume a spontaneous symmetry breaking due to a Higgs mechanism, but in view of the postulated broken symmetry of our model at the effective tree level, we need to account for the custodial symmetry to maintain $\rho=1$ at the tree level, in approximate agreement with the experimental data. The custodial symmetry gets violated by the hypercharge and by mass splitting within doublets. Because the mass splitting yields contributions proportional to the square of the masses, the top quark produces the dominant effect, $\Delta\rho_t\sim G_Fm_t^2$~\cite{Veltman1977}. On the other hand, the Higgs mass contribution is only logarithmic and proportional to the hypercharge coupling, $\Delta\rho_H\sim g_Y^2\ln(M_H/M_W)$. Top quark corrections with $M_H=0$ have been calculated up to three-loop order~\cite{Bij2001} and four-loop order~\cite{Chetyrkin2006}. The one-loop top quark radiative correction to $\rho$ is~\cite{Veltman1977}:
\begin{equation}
\Delta\rho_t=3x_t=3\frac{G_Fm_t^2}{8\sqrt{2}\pi^2}.
\end{equation}
Since the observed top quark mass is much larger than the $W$ and $Z$ masses, the higher-loop calculations can be performed by using an expansion in the external momenta. The Feynman integrals that have to be computed can be reduced to vacuum graphs which are easier to calculate. Known QCD corrections at the two-loop level are the leading $G_F^2m_t^4$ and sub-leading $g_{\rm weak}^2G_Fm_t^2$ contributions. The three-loop self-energy diagrams of the $W$ and $Z$ bosons also have to be accounted for.

The radiative correction $\Delta\rho$ due to the dominant top quark corrections can be cast in the form~\cite{Chetyrkin2006}:
\begin{equation}
\Delta\rho=3x_t\sum_{i=0}^3\biggl(\frac{\alpha_s}{\pi}\biggr)^i\Delta\rho_i,
\end{equation}
where $x_t=G_Fm_t^2/8\sqrt{2}\pi^2=0.0096$ and the factor 3 comes from the number of quark colors $n_c=3$. The renormalization method uses the $\overline{MS}$ scheme, $\alpha_s(m_t)=0.109$ and $\Delta\rho_0=1$. The result for the three-loop calculation is given by
\begin{equation}
\Delta\rho_3=-1.6799.
\end{equation}
Numerically, the two-loop and three-loop corrections are small compared to the one-loop correction~\cite{Chetyrkin2006}.

An important issue is whether the perturbation series is convergent, for perturbative QFT is only asymptotically convergent. Pad\'e approximation schemes have been used to investigate the convergence of the calculations but at present no clear and definite answer is available.

The Higgs boson contributions to the self-energy radiative corrections have been calculated up to three-loop order~\cite{Boughezal2005}. The three-loop Higgs corrections to the $\rho$ parameter read:
\begin{equation}
\rho_H=1+\Delta\rho^{(1)}_H+\Delta\rho^{(2)}_H+\Delta\rho^{(3)}_H.
\end{equation}
The results are given by
\begin{eqnarray}
\Delta\rho^{(1)}_H&=&-\frac{3}{4}\frac{g^2}{16\pi^2}\frac{s_w^2}{c_w^2}\ln\biggl(\frac{M_H^2}{M_W^2}\biggr),\\
\Delta\rho^{(2)}_H&=&\biggl(\frac{g^2}{16\pi^2}\biggr)^2\frac{s_w^2}{c_W^2}\frac{M_H^2}{M_W^2}(0.1499),\\
\Delta\rho^{(3)}_H&=&\biggl(\frac{g^2}{16\pi^2}\biggr)^3\frac{s_w^2}{c_w^2}\frac{M_H^4}{M_W^4}(-1.7282).
\label{deltarhores}
\end{eqnarray}
Here, we have
\begin{equation}
g^2=\frac{e^2}{s_w^2}=\frac{4\pi\alpha}{s_w^2}
\end{equation}
with $\alpha=1/137.036$ and $s_w^2=0.2312$.

For $M_H/M_W=2$ the results (110-112) give
\begin{equation}
\Delta\rho^{(1)}=-0.00078,\quad \Delta\rho^{(2)}=1.14\times 10^{-6},\quad \Delta\rho^{(3)}=-1.33\times 10^{-7}.
\end{equation}
The Tevatron search for the Higgs boson has excluded a Higgs mass in the range $158\,{\rm GeV}\leq M_H\leq 179\,{\rm GeV}$ at the $96\%$ confidence level, thereby excluding in this range a Higgs mass $M_H\sim 2M_W$. The $\Delta\rho^{(3)}$ becomes more important than $\Delta\rho^{(2)}$ for increasing values of $M_H$. The two contributions cancel each other at $M_H\sim 6M_W\sim 480$ GeV, and $\Delta\rho^{(3)}$ becomes equal to $\Delta\rho^{(1)}$ for $M_H\sim 2$ TeV. The motivation for performing the three-loop Higgs correction calculation was that a strongly coupled Higgs interaction could lead to effects mimicking the one-loop effects of a light Higgs boson~\cite{Boughezal2005}. The result of the calculation of $\Delta\rho^{(3)}$ shows that this is highly unlikely, for the sign of this correction is the same as the one-loop calculation of $\Delta\rho^{(1)}$. The three-loop calculation with an increasing Higgs mass only makes the $\Delta\rho_H$ grow faster, instead of permitting a partial cancelation of the one-loop correction. This makes the presence of a strongly interacting heavy Higgs very unlikely.

For the central value of $M_H=87$ Gev obtained from global fits to the EW data, we find the one-loop Higgs correction:
\begin{equation}
\Delta\rho_H^{(1)}=1.596\times 10^{-8},
\end{equation}
which is beyond the precision of current experiments. For $M_H=125$ GeV we get
\begin{equation}
\label{Higgs1}
\Delta\rho_H^{(1)}=5.015\times 10^{-4}.
\end{equation}
We can compare this result with the two-loop top quark correction~\cite{Bij2001}:
\begin{equation}
\Delta\rho^{(2)}_t=-2.218x_t^2-x_t^2\frac{M_Z^2}{m_t^2}\biggl(90.1-63\ln\biggl(\frac{M_Z^2}{m_t^2}\biggr)\biggr)
=-4.599\times 10^{-3},
\end{equation}
which is an order of magnitude larger. Thus, we have to compare the Higgs correction (\ref{Higgs1}) to the three-loop and four-loop top quark corrections to detect an appreciable contribution from the light Higgs sector. This Higgs correction would appear to be beyond the current experimental EW data detection.

We conclude that for the experimentally favored light Higgs, minimal standard EW model, our minimal EW model without a Higgs particle is consistent with current experimental data.

\section{Conclusions}

An electroweak model without a Higgs particle that possesses an intrinsic dynamically broken $SU_L(2)\times U_Y(1)$ has been developed, based on a finite UV complete quantum field theory. The theory is Poincar\'e invariant and unitary to all orders of perturbation theory and the symmetry breaking guarantees the photon has zero mass. We postulate that there never exists a massless $SU(2)\times U(1)$ symmetry phase. A fundamental energy scale $\Lambda_W$ enters into the calculations of the finite Feynman loop diagrams. A path integral is formulated that generates all the Feynman diagrams in the theory for the broken symmetry by postulating a path integral measure that is compatible with the broken symmetry at the quantum loop level. The self-energy boson loop graphs with internal fermions comprised of the observed 12 quarks and leptons are calculated to one-loop order and the radiative correction to the $\rho$ parameter is shown to be consistent to within errors with calculations of top quark dominated self-energy contributions without a light Higgs boson.

It is shown in a separate article~\cite{Moffat2008,Moffat2010} that the $W_LW_L\rightarrow W_LW_L$ and $e^+e^-\rightarrow W^+_LW^-_L$ amplitudes do not violate unitarity at the tree graph level due to the running with energy of the electroweak coupling constants $g(s), g'(s)$ and $e(s)$, which vanish sufficiently rapidly to cancel the $\sim s/M_W^2$ behavior for $\sqrt{s} > 1-2$ TeV. This is essential for the physical consistency of the model as is the case in the standard Higgs electroweak model.

The scattering amplitudes predicted by our model for $\Lambda_W > 1-2$ TeV will differ at high energies $\sqrt{s} > 1-2$ TeV compared to the standard Higgs model and this will allow the Higgsless and standard EW models to be distinguished from one another at the LHC.

There is no hierarchy problem in our Higgsless model, so the model does not require any new particles to be detected at the LHC to resolve this long-standing problem. The vector boson sector of our model does not possess a Landau pole~\cite{Moffat2010}. The fermion masses in our EW model are the measured masses and they and a phase fix the Cabibbo-Kobayashi-Maskawa (CKM) matrix. We do not attempt to explain the origin of the masses or the dynamical symmetry breaking of $SU(2)\times U(1)$. We can produce neutrino flavor mixing through a mass matrix with off-diagonal energy scales $\Lambda_{ff'}$.

\end{fmffile}

\section*{Acknowledgements}

I thank Viktor Toth, Denjoe McConnor and Bob Holdom for helpful and stimulating discussions. This work was supported by the Natural Sciences and Engineering Research Council of Canada. Research at the Perimeter Institute for Theoretical Physics is supported by the Government of Canada through NSERC and by the Province of Ontario through the Ministry of Research and Innovation (MRI).

\end{document}